\documentclass[runningheads]{llncs}
\usepackage[T1]{fontenc}
\usepackage[utf8]{inputenc}
\usepackage{graphicx}
\usepackage{amsmath,amssymb,amsfonts}
\usepackage{textcomp}
\usepackage{enumitem}
\usepackage{graphicx}
\usepackage{subfig}
\usepackage{algorithm}
\usepackage{algpseudocode}
\usepackage{booktabs}
\usepackage{array}
\usepackage{multirow}
\usepackage{pifont}
\usepackage[misc]{ifsym}
\usepackage{subfig}
\usepackage{url}
\usepackage{xurl}
\usepackage[hidelinks]{hyperref}

\usepackage{orcidlink}
\usepackage{comment}

\begin{document}
\title{A Collaborative Artificial Intelligence as a Service Composition Dataset}

\titlerunning{A Collaborative AIaaS Composition Dataset}

\author{
Deepak~Kanneganti\orcidlink{0009-0007-5860-2644}\textsuperscript{(\Letter)},
Sajib~Mistry\orcidlink{0000-0001-7513-3789},
Sheik~Fattah\orcidlink{0000-0002-9103-6089},
and Aneesh~Krishna\orcidlink{0000-0001-8637-5732}
}

\authorrunning{D. Kanneganti et al.}

\institute{
School of Electrical Engineering, Computing and Mathematical Sciences, \\Curtin University, Australia\\
\email{\{s.kanneganti,sajib.mistry,sheik.fattah,a.krishna\}@curtin.edu.au}
}

\maketitle            
\begin{abstract}
Artificial Intelligence as a Service (AIaaS) composition is an emerging research area that enables clients to combine multiple AI services to meet complex requirements. A recent extension of this paradigm is collaborative AIaaS composition, where multiple AI services are combined to create a unified solution. Research in this field requires datasets with service descriptions, composition requests, and corresponding composition solutions. However, no dataset contains all this information specifically for collaborative AIaaS composition. Existing datasets target traditional web services or sequential AI workflows and lack the AI-specific attributes required for realistic collaborative composition. To address these challenges, we present a collaborative AIaaS composition dataset containing 25,900 AIaaS services from multiple providers across 12 AI task families, along with 10,000 collaborative service requests. We further develop a Multi-Armed Bandit (MAB)-based collaborative composition algorithm that determines the composability of candidate service combinations prior to composition. Experimental results demonstrate that the dataset supports realistic collaborative AIaaS composition evaluation and related research in service recommendation, selection and QoS prediction. The dataset and implementation are publicly available at: \url{https://github.com/deepakkanneganti9/CAIaaS}
\end{abstract}
\keywords{Artificial Intelligence as a Service \and Service Composition\\ Composition Dataset \and Service Selection}
\vspace{-3mm}
\section{Introduction}

Artificial Intelligence as a Service (AIaaS) is a cloud-based service paradigm that provides infrastructure to build, train, and deploy AI models. Major providers such as \textit{Amazon}, \textit{Google}, and \textit{Hugging Face} offer a wide range of AIaaS solutions, covering tasks such as classification and predictive modeling. For example, \textit{Hugging Face}\footnote{\url{https://huggingface.co/docs/hub/main/en/index}} hosts more than \textit{two million} AI services and datasets through its platform. These services are widely adopted in data-driven applications across domains such as healthcare, industrial, and digital platforms~\cite{xie2024skyml}. Clients with heterogeneous service requirements often face challenges relying on a single AIaaS service~\cite{fu2023client}. For example, a healthcare environment may require human activity recognition (HAR) with 99\% accuracy across 100 activity classes, while also meeting a strict latency bound. Identifying a single service that satisfies all these requirements is often infeasible. This motivates the need for service composition.

\begin{figure*}[t]
  \centering
  \includegraphics[width=\linewidth]{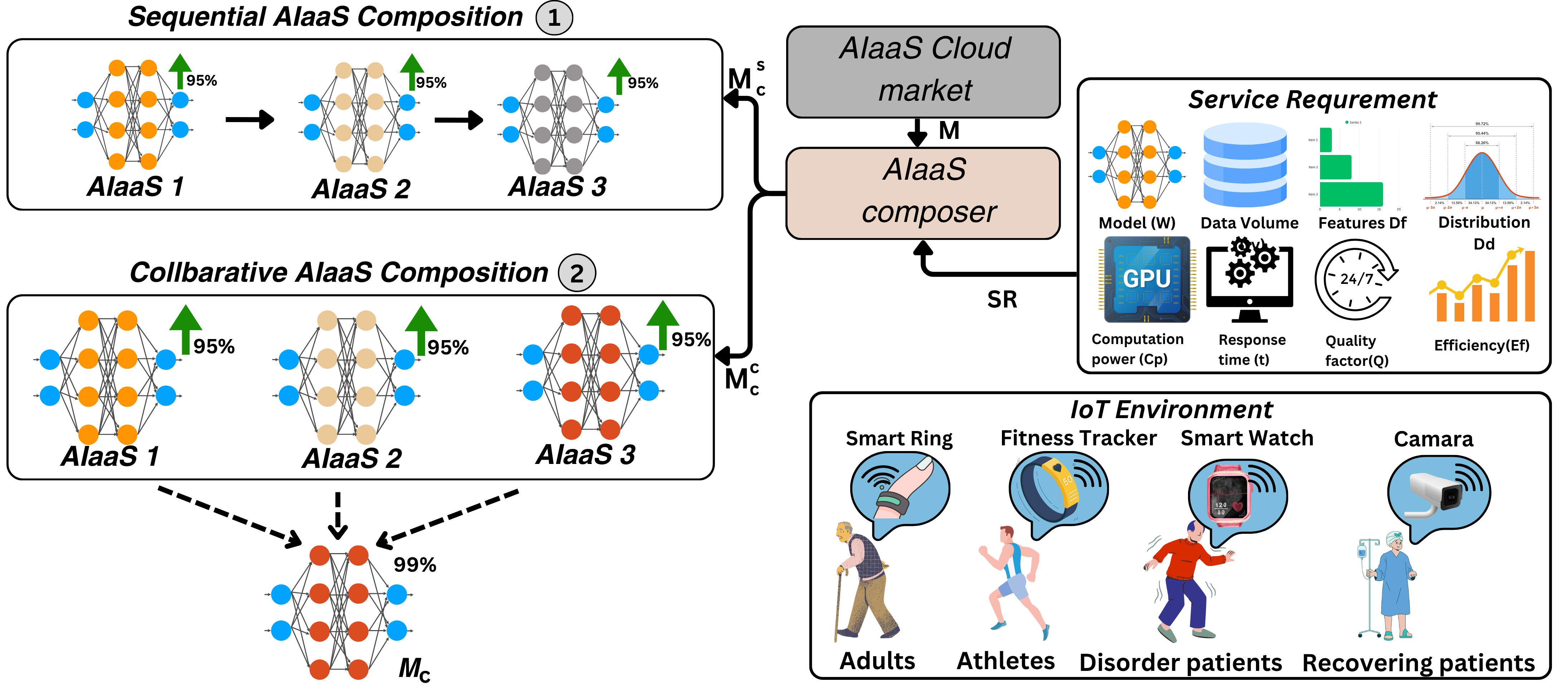}
  \caption{AIaaS Composition Dataset Motivation Scenario}
  \label{fig1}
\end{figure*}

AIaaS service composition is an emerging paradigm that integrates multiple AI services into a composite service to meet complex application requirements. It is broadly categorized into sequential and collaborative composition, as illustrated in Fig.~\ref{fig1}. In a sequential composition, services are organized in a workflow in which each service performs a specific task and passes its output to the next service~\cite{wang2024hsc}. For example, a healthcare application may convert a medical image into text and use it to generate a clinical summary for further analysis. In contrast, a collaborative composition combines multiple AIaaS services into a unified service, where each service contributes different functional capabilities or QoS characteristics~\cite{kanneganti2025adaptive}. For example, in a HAR application, one service can detect walking and running, while another service may recognize sitting activities. Composing these services results in a unified HAR service with broader functionality and improved overall performance.

Let us assume an AIaaS composition environment (\textit{see Fig.~\ref{fig1}}), where clients submit service requirements to an AIaaS composer to identify composition solutions. The composer performs tasks such as service selection, recommendation, and composition to meet these requirements. Building and evaluating such composition approaches requires datasets containing AIaaS service descriptions, service requirements, and corresponding composition solutions. However, identifying suitable composition solutions for different service requirements is computationally expensive and time-consuming due to the large composition search space. Furthermore, most AIaaS service information remains \textit{incomplete} and \textit{unclear}, as providers do not expose all model characteristics and QoS attributes needed for composition. To the best of our knowledge, there is no publicly available AIaaS dataset that provides this information. To address these limitations, we introduce the \textit{Collaborative AIaaS Service Composition Dataset}.

Existing studies in service computing have developed benchmark datasets such as QWS~\cite{al2008investigating}, WSDREAM~\cite{zheng2012investigating}, and OWLS-TC2~\cite{klusch2012overview} for traditional web and cloud services. These datasets lack the AIaaS-specific attributes required for composition, including training data specifications (e.g., modality and data distribution), model specifications (e.g., model type and parameters), and QoS attributes (e.g., evaluation metrics and resource usage). A recent study introduced the HSC dataset for large-scale AI service composition~\cite{wang2024hsc}. However, it primarily focuses on sequential composition and lacks these attributes, limiting its applicability to collaborative AIaaS composition.
To address these challenges, we introduce a comprehensive dataset for collaborative AIaaS composition research. Unlike existing datasets restricted to a single platform or a narrow set of models, the dataset includes heterogeneous AIaaS services integrating pre-trained and fine-tuned models from \textit{5 providers}. The dataset comprises \textit{25,900} AIaaS services spanning \textit{12 AI task} families, including classification, regression, detection, clustering, segmentation, and natural language processing (NLP). To reflect real-world usage, the framework incorporates a service requirement generator that models functional needs, QoS constraints, and user preferences to produce application-driven collaborative composition requirements. For each generated request, the framework derives composition solutions using a \textit{multi-armed bandit (MAB)-based AIaaS composability algorithm} to identify feasible and high-quality service combinations under varying conditions. The service requirements and corresponding composition solutions are evaluated using multiple optimization strategies to estimate baseline performance. This design maps service requirements to composition outcomes, enabling consistent comparison of composition techniques. The dataset therefore provides a benchmark for evaluating collaborative AIaaS composition solutions in diverse scenarios. Our key contributions are summarized as follows.
\begin{itemize}[itemsep=0ex, leftmargin=2ex]
\item We propose the \textit{collaborative AIaaS composition dataset} workflow designed to support collaborative AIaaS composition scenarios.
\item We construct a large-scale AIaaS Composition dataset containing \textit{25,900} AIaaS services from multiple providers evaluated on benchmark datasets.
\item We generate \textit{10,000} collaborative service requirements with diverse objectives, QoS constraints, preference weights, and corresponding composition solutions.
\item We develop a MAB-based AIaaS composability algorithm to generate benchmark composition solutions for evaluating AIaaS composition approaches.
\end{itemize}
\section{Related Work}
AIaaS composition has recently emerged as a new paradigm within the broader field of service composition~\cite{kanneganti2026machine,kanneganti2026performance}. Recent studies on AIaaS composition~\cite{kanneganti2025adaptive} and AIaaS federation~\cite{xie2024skyml} have demonstrated their potential to support advanced IoT applications. However, due to the lack of publicly available datasets and privacy constraints in AIaaS environments, most studies rely on individually developed AI models or manually collected service characteristics~\cite{fattah2020signature}. This leads to incomplete service information, limited composition diversity, and expensive evaluation processes for testing different composition scenarios~\cite{patel2024context}. Traditionally, service composition research has relied on datasets containing a limited number of fixed services. Early foundational resources, such as the QWS dataset, provided QoS measurements for 2,507 web services annotated across nine QoS parameters~\cite{al2008investigating}. Similarly, the WS-DREAM dataset later expanded the scale to 21,358 web services, enabling QoS analysis across dimensions such as response time, throughput, and failure probability~\cite{zheng2012investigating}. Another service composition work introduced two benchmark datasets: OWLS-TC, containing 1,083 services and 42 queries, and SAWSDL-TC, including 1,080 services and 42 requests, both annotated with multiple ontologies for semantic service retrieval evaluation ~\cite{klusch2012overview}. In addition, the Web Services Challenge (WSC) benchmark series supported the evaluation of functional and QoS-aware service composition using WSDL, OWL, and WSLA specifications ~\cite{blake2010wsc}.

Recently, the HSC dataset was introduced to support AI service composition research by providing 17,536 AI services across 20 AI tasks for sequential composition \cite{wang2024hsc}. Although the dataset significantly improves the scale of AI service composition research, it still has several limitations, including limited QoS attributes such as cost, insufficient internal AI parameters, and no support for collaborative AIaaS composition scenarios. Table~1 highlights the gaps in AIaaS-specific attributes and composability support, limiting the applicability of existing datasets to realistic AIaaS composition environments.

\begin{table}[!t]
\centering
\caption{Existing datasets for AIaaS service composition research.}
\label{tab1}
\scriptsize
\setlength{\tabcolsep}{2pt}
\renewcommand{\arraystretch}{1.05}
\resizebox{\columnwidth}{!}{%
\begin{tabular}{lccccc}
\toprule
\textbf{Datasets} &
\shortstack{\textbf{AIaaS}\\\textbf{specific}} &
\shortstack{\textbf{Functional}\\\textbf{attributes}} &
\shortstack{\textbf{QoS}\\\textbf{attributes}} &
\shortstack{\textbf{Internal}\\\textbf{parameters}} &
\shortstack{\textbf{Composability}\\\textbf{support}} \\
\midrule
QWS~\cite{al2008investigating}  & \ding{109} & \ding{109} & \ding{108} & \ding{109} & \ding{109} \\
WS-DREAM~\cite{zheng2012investigating}         & \ding{109} & \ding{109} & \ding{108} & \ding{109} & \ding{109} \\
OWLS-TC ~\cite{klusch2012overview}         & \ding{109} & \ding{108} & \ding{109} & \ding{109} & \ding{109} \\
IAIaaS~\cite{kanneganti2025adaptive} & \ding{108} & \ding{108} & \ding{119} & \ding{109} & \ding{119} \\
HSC~\cite{wang2024hsc}                & \ding{119} & \ding{108} & \ding{108} & \ding{119} & \ding{119} \\
\textbf{Collaborative AIaaS composition}     & \textbf{\ding{108}} & \textbf{\ding{108}} & \textbf{\ding{108}} & \textbf{\ding{108}} & \textbf{\ding{108}} \\
\bottomrule
\end{tabular}%
}
\vspace{-2mm}
\begin{flushleft}
\scriptsize \textbf{Note:} \ding{108} supported, \ding{119} partially supported, \ding{109} not supported.
\end{flushleft}
\end{table}
\vspace{-3mm}
\section{AIaaS Service Composition Dataset}
\begin{figure*}[t]
  \centering
  \includegraphics[width=\linewidth]{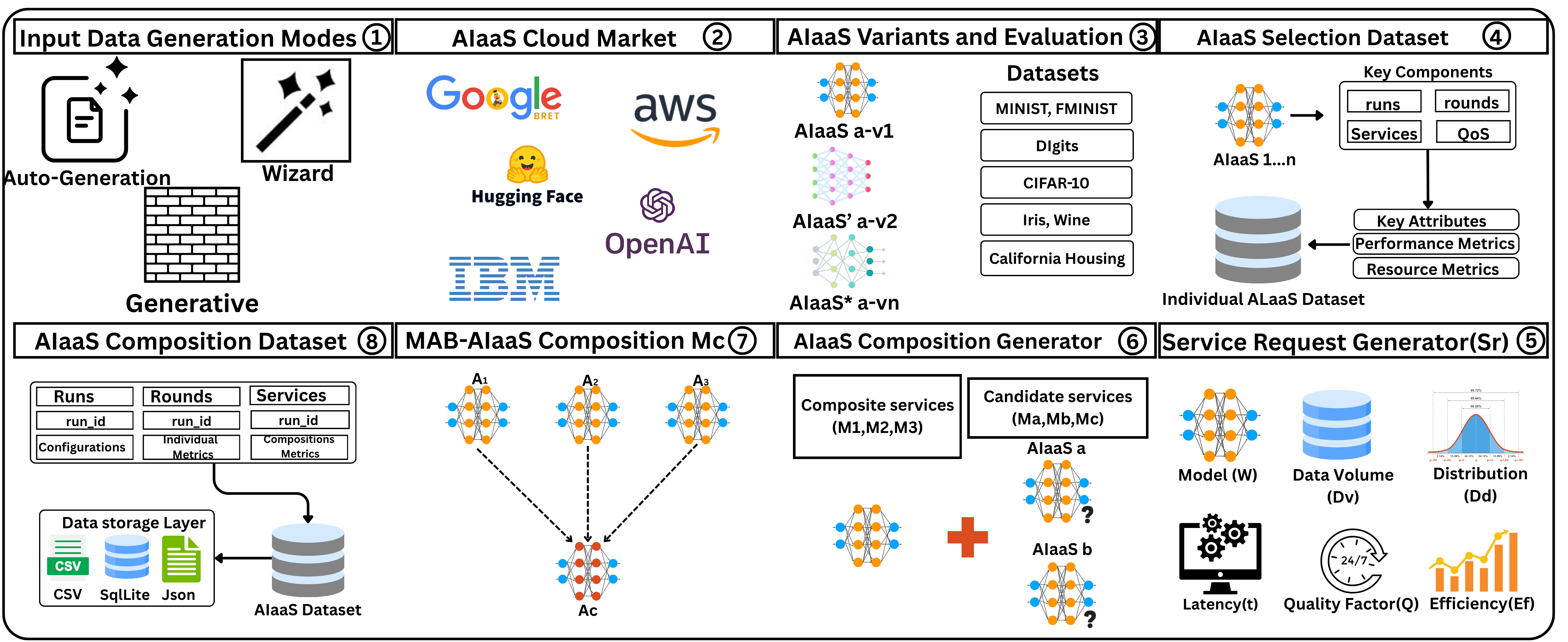}
  \caption{Overall workflow of the AIaaS service
composition dataset collection}
  \label{fig2}

\end{figure*}

In this section, we discuss the collaborative AIaaS composition dataset collection methodology, including AIaaS services, service requirements, composition scenarios, and AIaaS composition. Figure~\ref{fig2} illustrates the overall workflow of the collaborative AIaaS composition dataset collection. The workflow first collects heterogeneous inference-based and fine-tuned AIaaS services from multiple providers and extracts their attributes using benchmark datasets. It then generates realistic service requirements for collaborative composition scenarios. Finally, a MAB-based AIaaS composability algorithm is applied to generate composition solutions and construct the final collaborative AIaaS composition dataset for benchmarking and evaluation.

\subsection{AIaaS Service}
The rapidly growing AIaaS ecosystem offers diverse repositories and providers, enabling new applications through the composition of AIaaS services. To support realistic composition scenarios, we focus on services that are practically usable and composition-ready. We selected \textit{12 AI task} families spanning major domains such as natural language processing, computer vision, predictive modeling, clustering, regression, detection, and segmentation, as summarized in Table~\ref{tab2}. To reflect realistic AIaaS deployment scenarios, we construct services using both inference-ready and fine-tuned models. Since commercial AIaaS providers typically expose only limited model-level information to clients, fine-tuning enables the derivation of representative internal model characteristics and training behaviors of original services. To evaluate these services, we used widely adopted benchmark datasets associated with each task category to capture observable performance across functional and QoS attributes. On \textit{May 01, 2026}, we collected approximately \textit{1,500} top-ranked models across the selected tasks. These models were used as base models to construct multiple AIaaS service instances through different fine-tuning and configuration settings, resulting in more than \textit{25,900} services with diverse functional and performance characteristics, as shown in Table~\ref{tab2}. Each service was described using training data specifications, internal model parameters, and QoS-related attributes such as accuracy, latency, reliability, and cost efficiency, along with additional AIaaS-specific attributes. Table~\ref{tab3} summarizes the descriptions of key service attributes used in the dataset.
Unlike traditional dataset construction approaches that rely only on static metadata, we also evaluated runtime service behavior by repeatedly measuring response times under different interaction conditions to capture service-side variability. Finally, to ensure that functional and QoS attributes could be compared within a unified scale for downstream composition analysis, we applied min-max normalization to all applicable parameters, except those already bounded in the range of 0 to 1, such as reliability scores. As a result, the normalized dataset provides a consistent representation of service characteristics while preserving the relative structure of the original measurements.

\begin{table*}[!t]
\centering
\caption{AIaaS benchmark tasks, sources, datasets, and number of services}
\label{tab2}
\footnotesize
\renewcommand{\arraystretch}{1.08}
\setlength{\tabcolsep}{5pt}

\begin{tabular*}{\textwidth}{@{\extracolsep{\fill}}
    p{2.9cm}
    p{2.8cm}
    p{4.2cm}
    r@{}}
\noalign{\hrule height 0.8pt}
\textbf{Task Family} & \textbf{Source} & \textbf{Evaluation Dataset} & \textbf{\#} \\
\noalign{\hrule height 0.8pt}

Fill Mask & HF, BERT & WikiText-2 & 3,288 \\
Text Classification & HF, BERT & SST-2, AG News, IMDb & 3,008 \\
Tabular Regression & sklearn, XGBoost & California Housing & 2,998 \\
Sentence Similarity & HF, MiniLM & STS-B, MRPC & 2,377 \\
Image Classification & Keras, sklearn, ViT & CIFAR-10, MNIST, FMNIST & 3,978 \\
Object Detection & HF, DETR & COCO & 2,521 \\
Token Classification & HF, BERT & CoNLL-2003, WNUT-17 & 2,045 \\
Image Segmentation & HF, SegFormer & ADE20K & 1,174 \\
Clustering & sklearn, KMeans & Iris, Wine & 1,000 \\
Regression & sklearn & Diabetes, California House & 1,000 \\
Text Generation & HF, GPT-2, T5 & WikiText-2 & 2,011 \\
Text-to-Text Gen & HF, T5 & XSum & 500 \\
\noalign{\hrule height 0.8pt}
\textbf{Total} & \multicolumn{2}{c}{} & \textbf{25,900} \\
\noalign{\hrule height 0.8pt}
\end{tabular*}
\end{table*}
\vspace{-5mm}
\begin{table*}[!t]
\centering
\caption{Formal AIaaS service attributes, symbols, and descriptions}
\label{tab3}
\footnotesize
\renewcommand{\arraystretch}{1.08}
\setlength{\tabcolsep}{5pt}
\begin{tabular*}{\textwidth}{@{\extracolsep{\fill}} p{3cm} c p{10cm}}
\noalign{\hrule height 0.8pt}
\textbf{Attribute} & \textbf{Symbol} & \textbf{Description} \\
\noalign{\hrule height 0.8pt}

\textit{Data distribution}    & $\Delta$ & Data distribution type, such as IID or non-IID. \\
\textit{Dataset size}         & $D_n$    & Number of training samples. \\
\textit{Model parameters}     & $w$      & Model weights and bias. \\
\textit{Accuracy}             & $A$      & Final task performance score. \\
\textit{Latency}              & $L$      & Average inference response time. \\
\textit{Reliability score}    & $R$      & Consistency of service performance. \\
\textit{Compute time}    & $T$      & Average computation time. \\
\textit{Resource cost score}  & $C$      & Operational resource cost. \\
\textit{Cost efficiency}      & $CE$     & Performance per unit cost. \\
\textit{Model size}           & $M$      & Number of model parameters. \\
\noalign{\hrule height 0.8pt}
\vspace{-5mm}
\end{tabular*}
\end{table*}

\begin{figure*}[t]
  \centering
  \includegraphics[
      width=0.99\textwidth,
      height=0.25\textheight,
      keepaspectratio
  ]{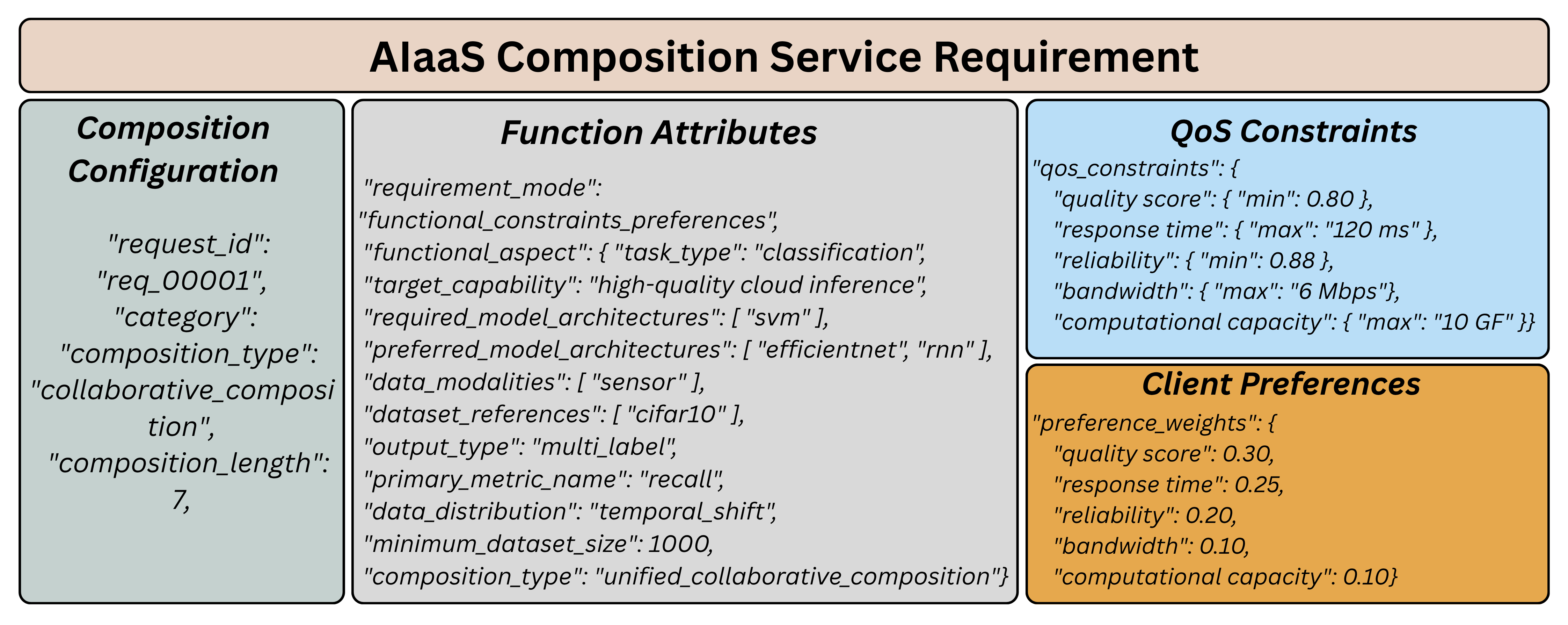}
  \caption{An example of an AIaaS Composition Service Requirement}
  \label{fig3}
\end{figure*}
\vspace{-6mm}
\subsection{Service Requirement}
Service requests are a key aspect of service composition, where clients define
functional and QoS requirements, preference weights, and constraints. Due to
the difficulty of collecting real-world AIaaS requests at scale, we generate
requests based on established artificial intelligence selection principles
from prior studies~\cite{fu2023client}. These principles reflect
application-driven preferences in AIaaS systems, including QoS prioritization.
For example, healthcare applications prioritize high accuracy and low latency,
whereas financial applications emphasize reliability and resource efficiency.
Each principle defines a base preference weight vector over the QoS attributes.
To generate a request, we apply a bounded perturbation to this vector and
renormalize the weights to sum to one. If the perturbation changes the defined
attribute ordering, the request is discarded and resampled. This preserves
attribute priorities while producing varied, non-identical requests.
Figure~\ref{fig3} illustrates a fundamental AIaaS service request structure.

\begin{algorithm}[!b]
\caption{AIaaS Service Requirement Generation}
\label{alg1}
\small
\begin{algorithmic}[1]
\Require Dataset $MD$, profiles $R$, principles $P$, requests $N$
\Ensure Service requests $SR$
\State $SR \gets \emptyset$
\For{$i = 1$ to $N$}
    \State Select profile $r_i \in R$ and principle $p_i \in P$
    \State Extract $F_i$, $QoS_i$, $PW_i$, and $C_i$ from $r_i$
    \Repeat
        \State Perturb $PW_i$ within bounds of $p_i$
    \Until{attribute order of $PW_i$ matches $p_i$}
    \State Randomly select attributes from $F_i \cup QoS_i \cup C_i$
    \ForAll{selected attribute $a$}
        \State Apply principle $p_i$ and update $a$ within valid bounds
    \EndFor
    \State Derive collaborative size $K_i$ from $MD$
    \State $sr_i \gets \langle F_i, QoS_i, PW_i, C_i, K_i \rangle$
    \State $SR \gets SR \cup \{sr_i\}$
\EndFor
\State \Return $SR$
\end{algorithmic}
\end{algorithm}

We design an AIaaS service requirement generation algorithm (\textit{SR}) for
collaborative composition (Algorithm~1). The algorithm constructs requests by
selecting preference profiles and AI service selection principles from the AIaaS
dataset. Each generated request includes functional requirements ($F$), QoS
requirements ($QoS$), preference weights ($PW$), and constraints ($C$), which
are derived using this perturb-and-reject procedure to reflect realistic
AIaaS selection behavior. We manually inspected a random subset of generated
requests against known AIaaS deployment patterns reported in prior
work~\cite{fu2023client,kanneganti2025adaptive}. The assigned priorities
aligned with expected behavior for the corresponding application domain. The
algorithm then determines the collaborative service group size for each
request, producing a complete service request specification ready for
composition solving.

\begin{table*}[t]
\centering
\caption{AIaaS composability function for collaborative composition}
\label{tab:aiaas_composability}
\scriptsize
\setlength{\tabcolsep}{4pt}
\renewcommand{\arraystretch}{1.2}

\begin{tabular}{|p{4cm}| p{7cm}|}
\noalign{\hrule height 0.8pt}
\textbf{Attributes} &
\textbf{Collaborative composition ($C_c$)}
\\
\noalign{\hrule height 0.8pt}

Data parameters ($\Delta, D_n$) &
$\displaystyle
1-\frac{1}{m}\sum_{j=1}^{m}
\frac{
\max_{i=1}^{k}(\Delta_{ij})
-
\min_{i=1}^{k}(\Delta_{ij})
}
{
\sum_{i=1}^{k}\Delta_{ij}
}
$
\\\hline

Model parameters ($w$, Loss) &
$\displaystyle
\frac{1}{k}\sum_{i=1}^{k}
\frac{
\langle \nabla w_i,\nabla w_c\rangle
}
{
\|\nabla w_i\|\|\nabla w_c\|
}
$
\\\hline

Computation time ($T$) &
$\displaystyle
1-\frac{1}{k}\sum_{i=1}^{k}|T_i-\mu_T|
$
\\\hline

Latency ($L$) &
$\displaystyle
\min\left(
1,
\frac{L_{\max}}
{\sum_{i=1}^{k}L_i}
\right)^{\alpha}
$
\\\hline

Accuracy ($A$) &
$\displaystyle
1-\frac{1}{k}\sum_{i=1}^{k}|A_i-\mu_A|
$
\\\hline

Reliability score ($R$) &
$\displaystyle
1-\frac{1}{k}\sum_{i=1}^{k}|R_i-\mu_R|
$
\\\hline

Cost ($C$) &
$\displaystyle
1-\frac{1}{k}\sum_{i=1}^{k}C_i
$
\\\hline

Objective function &
$\displaystyle
CC=\sum_{j=1}^{m}\lambda_w^j C_c^j
$
\\\hline

\noalign{\hrule height 0.8pt}
\end{tabular}
\end{table*}
\vspace{-3mm}
\subsection{AIaaS Composition Solution} 
To evaluate an AIaaS composite solution, we define the overall composability objective function for collaborative composition over a candidate service combination $S_k=\{M_1,M_2,\dots,M_k\}$ as follows:

\begin{equation}
\max \; CC(S_k)=\sum_{j \in \{1,2,..7\}} \lambda_w^j C_c^j(S_k)
\label{eq:aiaas_obj}
\end{equation}

where $C_c^j(S_k)$ denotes the $j$th collaborative composability score over the service combination $S_k$, and $\lambda_w^j$ represents the corresponding preference weight defined by the service requirement. Collaborative composition evaluates the compatibility and interaction behavior among multiple services. The detailed collaborative composability formulations are presented in Table~\ref{tab:aiaas_composability}. The formulation is designed to capture the composability characteristics identified in existing studies on AI service and collaborative learning environments, where both functional and QoS attributes influence the effectiveness of service composition \cite{fu2023client,kanneganti2025adaptive}. Functional characteristics such as data heterogeneity, data volume, model parameter alignment, and training loss behavior are considered to evaluate learning compatibility among AIaaS services. In addition, QoS-related attributes including computation time, latency, accuracy, reliability, and cost are incorporated to assess the operational quality and deployment feasibility of the composed solution. The formulation also supports application-driven service selection through preference weights, allowing different functional and QoS criteria to be prioritized according to service requirements and deployment objectives.

To ensure that the selected service combination satisfies the requested constraints, a penalty term is incorporated to capture deviations from the target functional and QoS requirements. Let $O(S_k)$ denote $CC(S_k)$. The penalty-aware composability objective is then defined as:

\begin{equation}
\max \; \hat{O}(S_k)
=
O(S_k)
-
\sum_{j \in \mathcal{R}}
\eta_j
\left|q_j(S_k)-q_j^{\,r}\right|^2
\label{eq:penalty_obj}
\end{equation}

where $\mathcal{R}$ denotes the set of constrained service request attributes, $q_j(S_k)$ represents the aggregated value of the $j$th attribute achieved by the composed AIaaS solution, $q_j^{\,r}$ denotes the corresponding target value specified in the service request, and $\eta_j$ represents the penalty coefficient associated with the $j$th constraint. The penalty term reduces the overall composability objective when the selected service combination deviates from the requested functional and QoS requirements. To evaluate AIaaS composition strategies, the objective is to identify service compositions that achieve optimal composability objective values, where a higher value of $\hat{O}(S_k)$ indicates a better composition. 


\begin{algorithm}[!b]
\caption{Multi-Armed Bandit based AIaaS Composition}
\label{alg2}
\begin{algorithmic}[1]
\State \textbf{Input:} $\langle SR, MD, SA, K, train\rangle$ 
\State \textbf{Output:} $bestS_k, bestCS$ 
\hfill $\triangleright$ Best service combination and best composability score

\State Generate candidate service combinations $\Omega$ from $MD$ and $SA$ based on $SR$ with size $K$
\State Initialize $r_{\text{est}}(S_k) \gets 0,\; n(S_k) \gets 0$ for all $S_k \in \Omega$
\State $bestCS \leftarrow -\infty,\; bestS_k \leftarrow$ None

\For{$t = 1$ to $train$}
    \ForAll{$S_k \in \Omega$}
        \State $ucb(S_k) \leftarrow 
        \begin{cases}
        \infty, & n(S_k)=0\\
        r_{\text{est}}(S_k)+\sqrt{\frac{2\log t}{n(S_k)}}, & otherwise
        \end{cases}$
    \EndFor
    \State $S_k^{*} \leftarrow \arg\max_{S_k \in \Omega} ucb(S_k)$
    \State $cs \leftarrow CC(S_k^{*})$
    \hfill $\triangleright$ Collaborative composition
    \State $n(S_k^{*}) \leftarrow n(S_k^{*}) + 1$
    \State $r_{\text{est}}(S_k^{*}) \leftarrow r_{\text{est}}(S_k^{*}) + 
    \frac{cs-r_{\text{est}}(S_k^{*})}{n(S_k^{*})}$
    \If{$cs > bestCS$}
        \State $bestCS \leftarrow cs,\; bestS_k \leftarrow S_k^{*}$
    \EndIf
\EndFor
\State \Return $bestS_k, bestCS$
\end{algorithmic}
\end{algorithm}
The composition process inherently forms a combinatorial optimization problem, as multiple services with varying capabilities must be selected and combined to satisfy given requirements. For instance, if the composition length is \( k = 2 \) and the total number of available services is \( n = 100 \), the number of possible combinations is \( C(n, k) = \frac{n!}{k!(n-k)!} = 4950 \), illustrating how rapidly the search space expands and making exhaustive exploration computationally infeasible. To address this challenge, candidate services are first filtered based on the functional and QoS requirements of each request, reducing the search space before optimization. Specifically, a multi-armed bandit (Algorithm~\ref{alg2}) is used for relatively smaller search spaces, enabling efficient exploration--exploitation trade-offs. For larger and more complex search spaces, a genetic algorithm with increased population size and iterations is adopted to approximate near-optimal compositions. These optimization strategies enable the discovery of high-quality AIaaS compositions that serve as benchmarks for evaluating different collaborative composition approaches.

\vspace{-3mm}

\section{Statistical Analysis} 

\begin{figure}[!b]
    \centering
    \subfloat[]{%
        \includegraphics[
            width=.49\textwidth
        ]{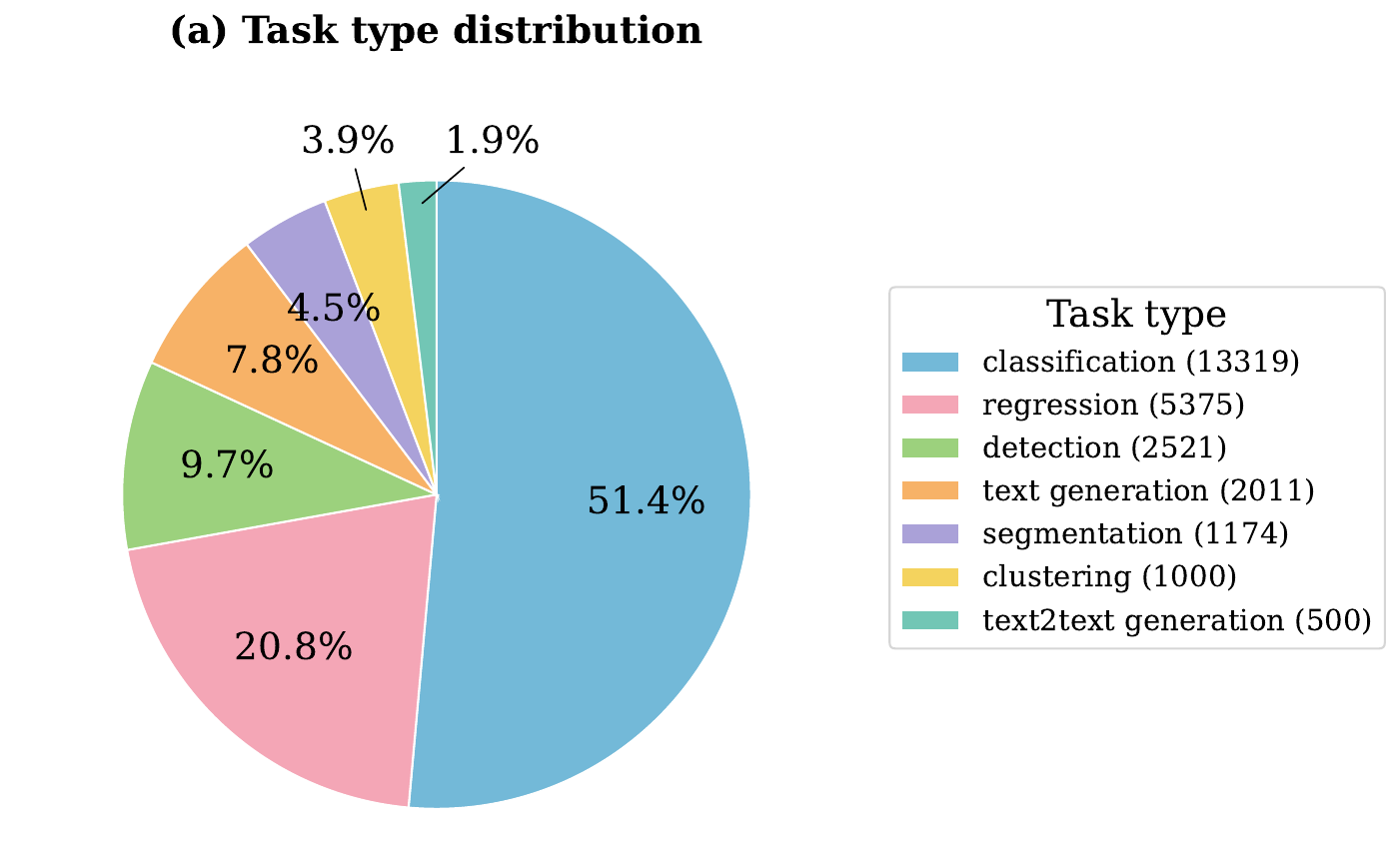}%
    }
    \hfill
    \subfloat[]{%
        \includegraphics[
            width=.49\textwidth
        ]{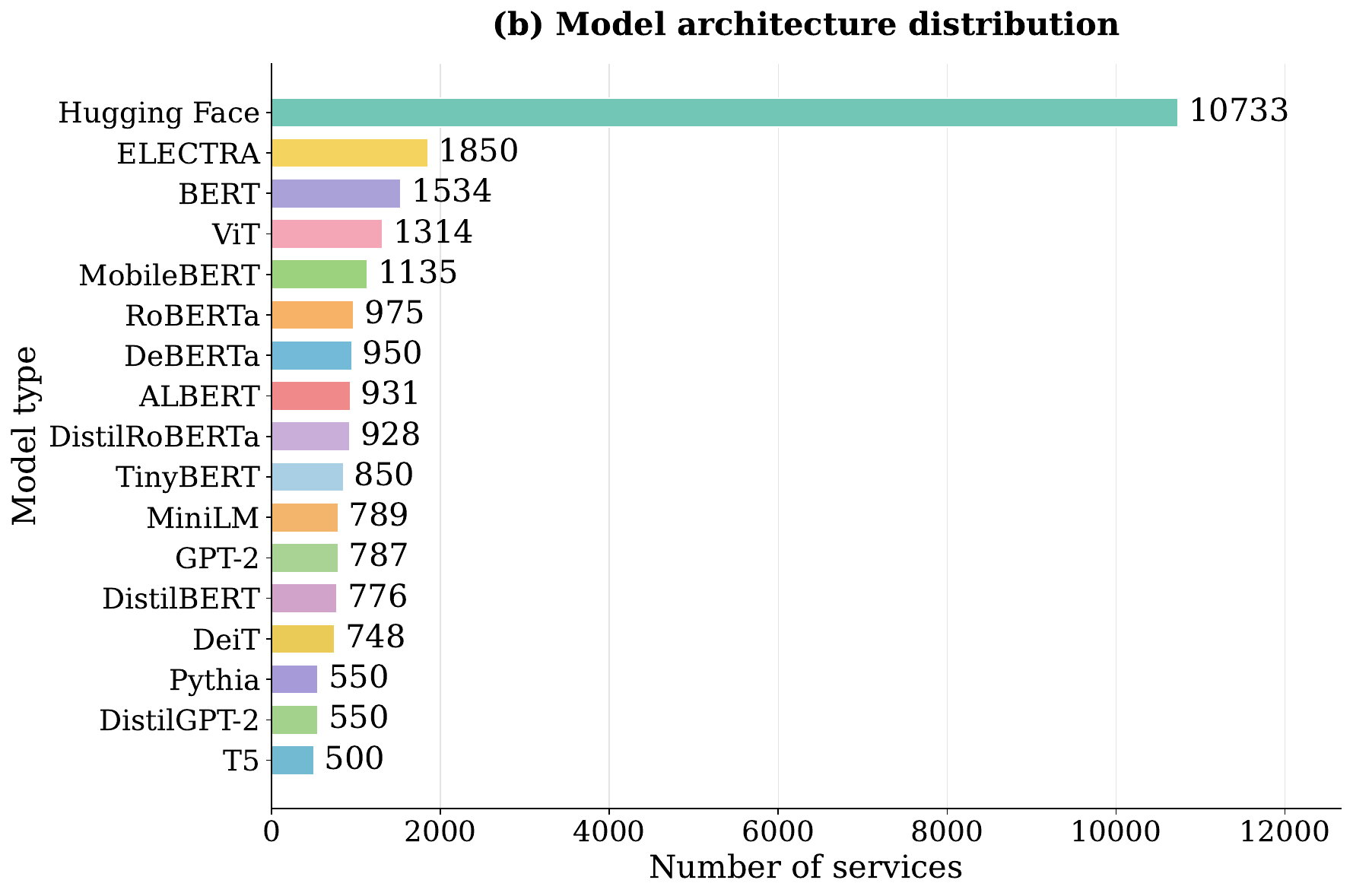}%
    }\\[0.5em]
    \subfloat[]{%
        \includegraphics[
            width=.49\textwidth
        ]{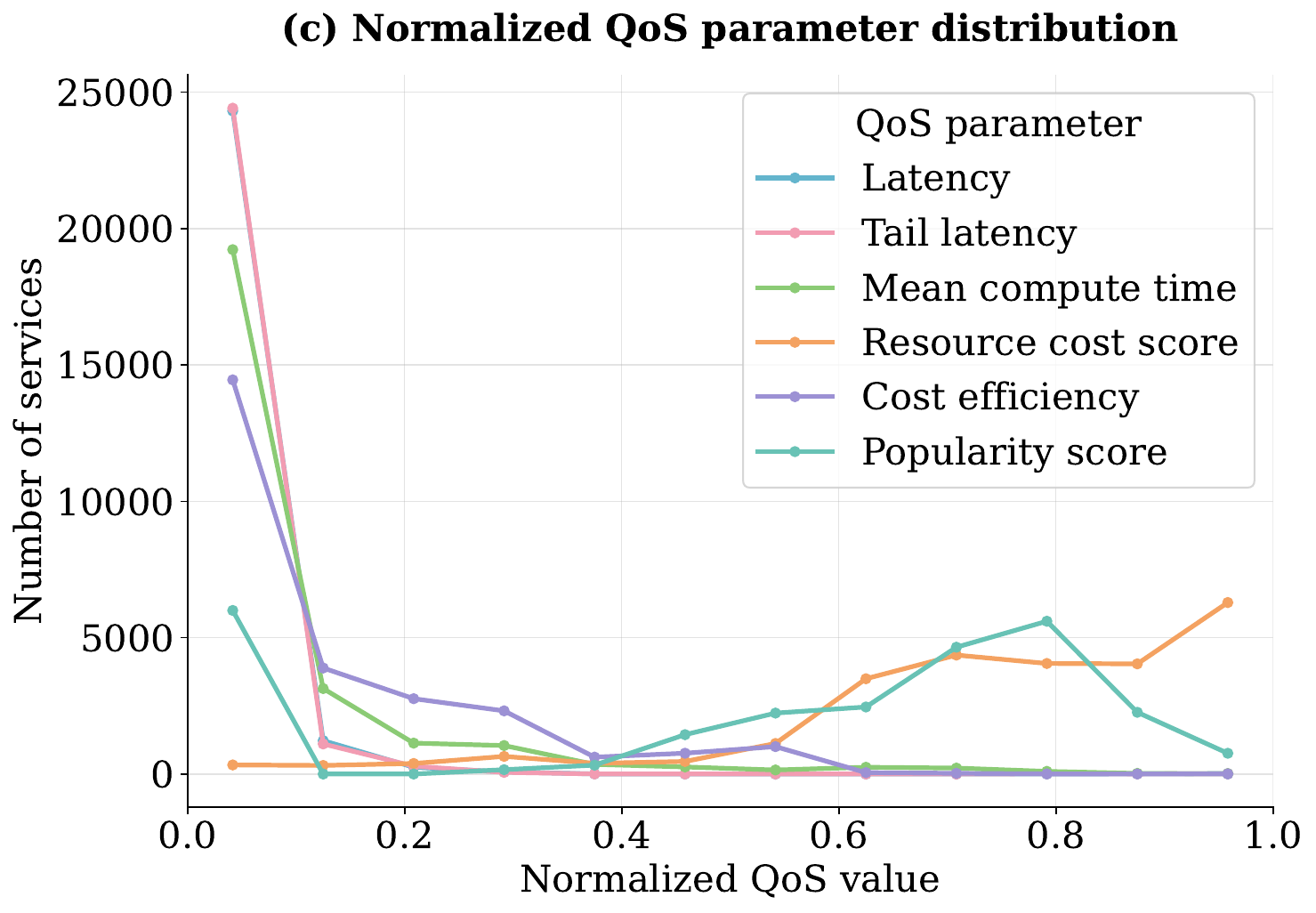}%
    }
    \hfill
    \subfloat[]{%
        \includegraphics[
            width=.49\textwidth
        ]{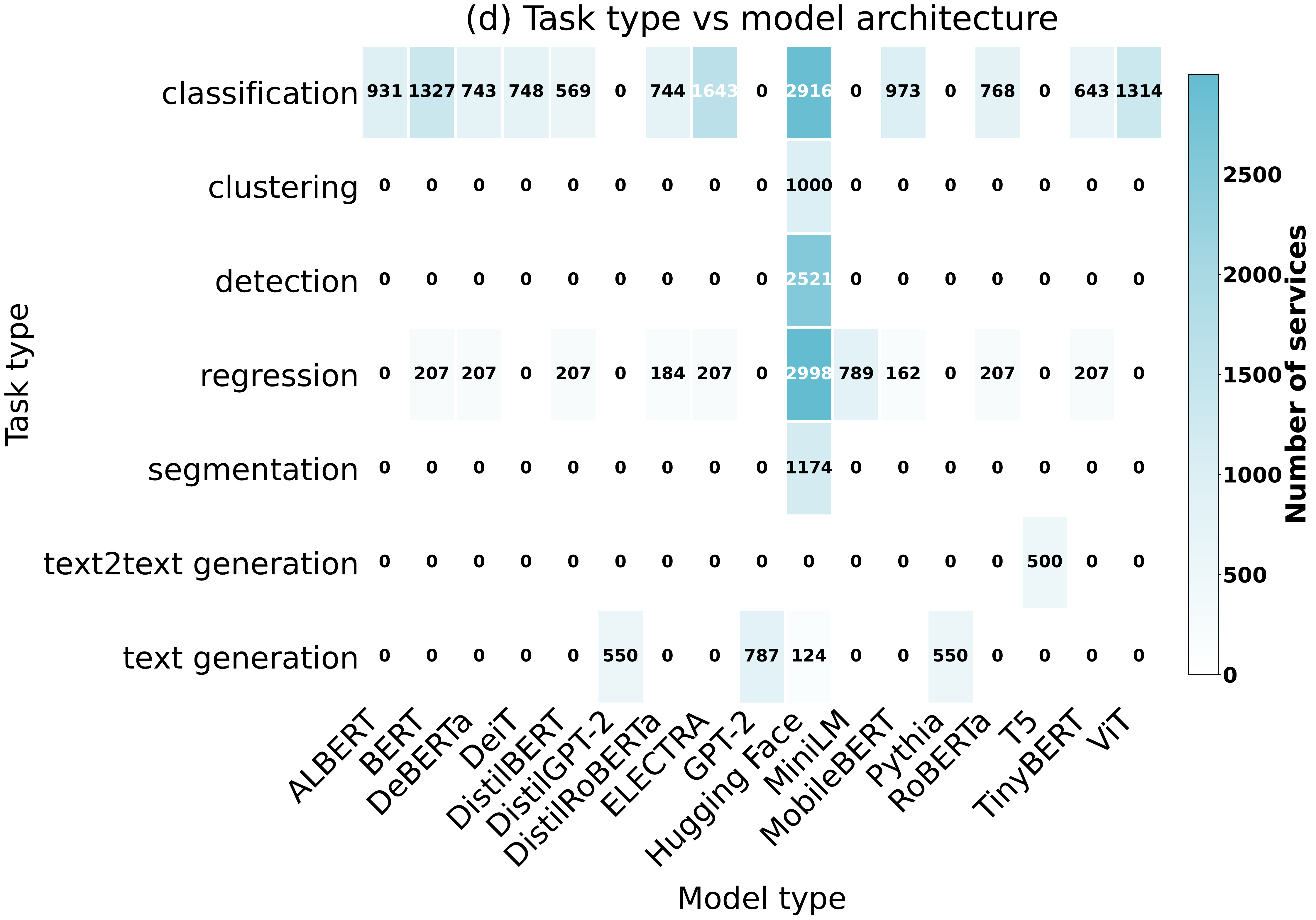}%
    }
    \caption{Statistical analysis of the AIaaS composition dataset showing:
    (a) task type distribution,
    (b) model architecture distribution,
    (c) normalized QoS parameter distribution, and
    (d) task type versus model architecture associations.}
    \label{fig4}
    \vspace{-5mm}
\end{figure}
The collaborative AIaaS composition dataset comprises 25,900 publicly available AI services with their corresponding functional and QoS attributes, making it one of the largest publicly accessible datasets for AIaaS composition research. Figure~\ref{fig4} presents a statistical analysis of the generated AI services across task families, model architectures, QoS attributes, and task--architecture relationships. Figure~\ref{fig4}a illustrates the distribution of major AI task families, including classification, regression, detection, clustering, segmentation, and text-generation tasks. Classification dominates the dataset with more than 13,000 services (51.4\%), while the remaining task families provide diversity for heterogeneous composition scenarios. Figure~\ref{fig4}b presents the distribution of model architectures. Transformer-based architectures, particularly Hugging Face and BERT-related variants, constitute a major portion of the dataset, alongside other architectures such as RoBERTa, ELECTRA, GPT-based models, and traditional AI approaches. This demonstrates the architectural diversity of the collected AI services. Figure~\ref{fig4}c shows the normalized QoS parameter distributions. Since QoS attributes are measured at different numerical scales, min--max normalization is applied to project all attributes into a unified range. The resulting distributions reveal variations and trade-offs among latency, and cost efficiency. Finally, Figure~\ref{fig4}d illustrates the relationship between task families and model architectures. The heatmap shows that transformer-based architectures are strongly associated with classification and text-oriented tasks, while other task families exhibit comparatively smaller architectural distributions. Overall, these observations show that the dataset captures diverse functional, architectural, and QoS characteristics required for evaluating AIaaS composition approaches.
\vspace{-3mm}
\section{Benchmark}
To further advance research on AIaaS service composition, we develop a benchmark framework for collaborative composition using the collaborative AIaaS composition dataset. The benchmark is designed to support researchers in accurately evaluating and comparing composition solutions.
\vspace{-4mm}
\subsection{Experiment Design}
\subsubsection{AIaaS Composition Scenario Setup.} The collaborative AIaaS composition dataset comprises 10,000 collaborative AIaaS service composition cases. The generated composition cases are organized based on different composition sizes, service request configurations, and composability objective functions to support diverse evaluation settings across collaborative composition environments.
\vspace{-2mm}
\subsubsection{Baseline techniques.}
In this section, we evaluate the average solution quality of benchmark service composition techniques using the collaborative AIaaS composition dataset. We measure the average solution quality of AIaaS compositions using the MAB-based AIaaS composability algorithm solutions for given service requirements. We consider our proposed approach as the baseline and compare its performance with recent service composition techniques that have been widely adopted over the last five years. These include traditional and metaheuristic-based techniques such as Genetic Algorithm, Epsilon-Greedy, Greedy, and Random Search. In addition, we consider several state-of-the-art optimization approaches, including DAAGA~\cite{yang2019dynamic}, MWOA~\cite{jin2022eagle}, CSSA~\cite{li2022novel}, SDFGA~\cite{li2020sdf}, BPSC-GA~\cite{xu2020domain}, and PK-IDPSO~\cite{wang2024particle}, which utilize historical service composition information during optimization. We then measure the average solution quality of AIaaS compositions generated using the benchmark composability model to evaluate the quality of the composition solutions provided with the dataset. All experiments were conducted on an Intel Core i7 machine with 16\,GB of RAM using Python. The source code is publicly available in the repository\footnote{\url{https://github.com/deepakkanneganti9/CAIaaS}}.



\begin{figure*}[t]
  \centering
  \includegraphics[
      width=.99\textwidth,
      height=0.25\textheight,
      keepaspectratio
  ]{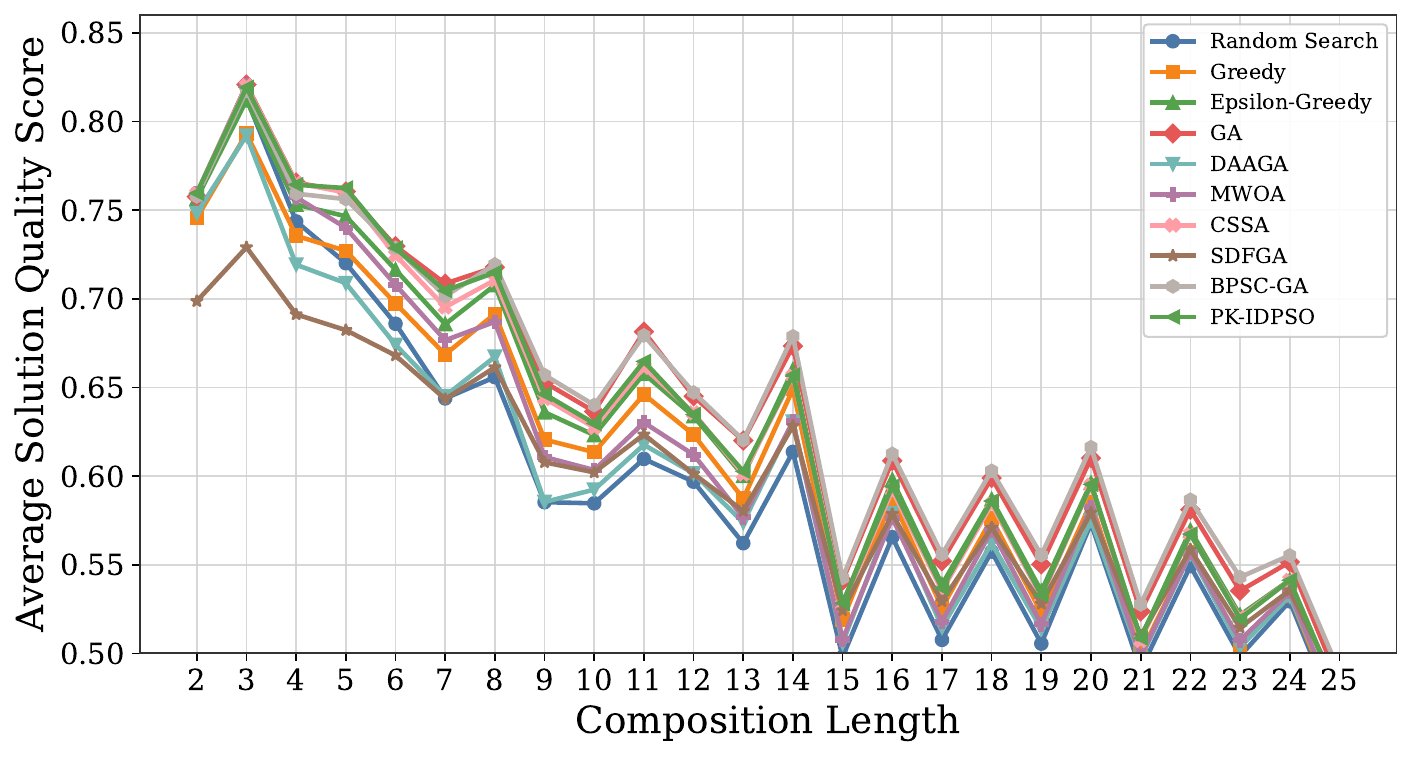}
  \caption{Average service quality score across different composition lengths.}
  \label{fig6}
\end{figure*}
\vspace{-2mm}
\subsection{Evaluation Metrics.} The performance of collaborative AIaaS composition strategies is evaluated using the average service composition quality score, defined as:
\begin{equation}
ASQ(sol)=\frac{\hat{O}(sol)}{\hat{O}(sol^{*})}
\label{eq:composition_quality}
\end{equation}
where $sol$ denotes the generated composition solution and $sol^{*}$ represents the highest-quality composition solution identified for the corresponding service request. A higher composition quality score indicates a better composition outcome, with values closer to 1 reflecting stronger alignment with the target functional and QoS requirements. To ensure a fair comparison, all optimization approaches are evaluated under identical experimental settings and composition constraints. The proposed MAB-based composability algorithm is used as the reference method for identifying $sol^*$ in Eq.~3. Since exhaustive evaluation of all possible service combinations is computationally infeasible for the collaborative AIaaS composition dataset due to the combinatorial growth discussed in Section~3.3, the true global optimum cannot be determined. Therefore, the solution produced by the MAB-based approach is treated as the best available reference for each service request rather than a verified global optimum. Accordingly, the $ASQ$ scores presented in Fig.~5 quantify the relative composition quality achieved by each optimization technique with respect to this reference. This evaluation reflects the practical objective of the proposed MAB-based approach, which is to provide a scalable and effective composition solution for large-scale environments where exhaustive verification is not feasible.

\vspace{-3mm}
\subsection{Experimental Results}
We evaluate the proposed MAB-based AIaaS composition algorithm against different optimization approaches under collaborative composition scenarios. Figure~\ref{fig6} shows that traditional approaches such as Random Search and Greedy exhibit lower and less stable performance for longer service combinations, demonstrating their limited applicability in large-scale AIaaS composition environments. In contrast, metaheuristic-based approaches including GA, DAAGA, MWOA, CSSA, SDFGA, BPSC-GA, and PK-IDPSO generally produce better composition solutions, highlighting the importance of optimization-based exploration in complex composition spaces. The results further demonstrate that the proposed MAB-based composability algorithm provides an effective benchmarking mechanism for evaluating AIaaS composition techniques. Moreover, the relatively small performance gaps across different optimization approaches indicate that the diverse functional and QoS attributes of the collaborative AIaaS composition dataset support consistently feasible composition solutions across varying service combinations.


\vspace{-2mm}
\section{Discussion and Research Implications}
The proposed collaborative AIaaS composition dataset supports several key tasks in service computing. In this section, we discuss problems beyond AIaaS composition, including AIaaS service recommendation, service selection and composition, service classification, and QoS prediction. We also highlight current limitations and open challenges.
\vspace{-2mm}
\subsection{AIaaS Service Recommendation}
AIaaS service recommendation focuses on identifying services that best match user requirements by analyzing functional needs and preference patterns. User requirements are often expressed in natural language or abstract constraints, requiring mapping to service attributes. The proposed \textit{collaborative AIaaS composition dataset} supports this task by providing thousands of mappings between requirements and services, covering diverse combinations of model type, data modality, and QoS preferences. These mappings enable researchers to evaluate how effectively service recommendation approaches identify suitable services under heterogeneous conditions, where services differ in capability, specialization, and performance tradeoffs.

\vspace{-2mm}
\subsection{AIaaS Service Selection and Composition}
AIaaS service selection and composition aim to identify and combine multiple services to satisfy complex requirements while considering both functional compatibility and QoS constraints. Unlike recommendation, composition explicitly models inter-service dependencies, workflow feasibility, and overall system performance. The collaborative AIaaS composition dataset is designed to support this problem by generating \textit{composition-aware service representations}. Each requirement instance is associated with candidate services and feasible compositions derived from functional compatibility (e.g., input-output alignment, data specifications) and QoS attributes (e.g., accuracy, latency, reliability). The dataset incorporates diverse service types, including pre-trained, fine-tuned, and hybrid models, enabling evaluation under realistic conditions. The structured generation of requirement--composition pairs also allows analysis of trade-offs between solution optimality and feasibility.
\vspace{-2mm}
\subsection{AIaaS Service Classification}
Service classification categorizes AIaaS services based on their functional descriptions and attributes, improving service discovery, selection, and composition efficiency. The collaborative AIaaS composition dataset includes diverse services spanning NLP, computer vision, predictive modeling, and clustering. Each service is described using functional attributes (e.g., model type, data modality, output structure) and QoS characteristics. The scale and diversity of pre-trained and fine-tuned services provide a strong benchmark for classification tasks while reflecting realistic challenges such as uneven category distributions and overlapping functionalities.
\vspace{-2mm}
\subsection{QoS Prediction in AIaaS}
QoS prediction aims to estimate unknown or future service performance metrics based on observed attributes. This is critical for enabling reliable service selection and composition under uncertainty. The collaborative AIaaS composition dataset supports QoS prediction by providing multiple QoS attributes for each service, including performance and system-level indicators. These attributes are derived from both platform metadata and empirical evaluation. The normalized QoS parameters ensure consistency across services, enabling the development and validation of predictive models. The dataset also supports analysis of trade-offs between QoS attributes and functional requirements.
\vspace{-3mm}
\section{Limitations and Threats to Validity}
The collaborative AIaaS composition dataset and workflow may be subject to several threats to validity. First, the dataset is constructed using publicly available AIaaS services, model repositories, and benchmark datasets. As a result, some service attributes and metadata may be unavailable due to limitations in publicly advertised specifications. In addition, the fine-tuned services in the dataset were generated by further training publicly available pre-trained inference models under controlled evaluation settings. Therefore, the generated services may not fully capture all deployment-specific optimizations used in commercial AIaaS environments. However, the objective of the dataset is not to provide absolute judgments on individual AI services, but to offer a realistic and reproducible benchmarking environment for evaluating AIaaS composition strategies under heterogeneous service and composition settings. To improve reproducibility and reduce experimental bias, the complete dataset generation workflow, evaluation configurations, and service extraction parameters are publicly available through the provided GitHub repository. The framework also allows researchers to customize evaluation settings, extend service attributes, and generate additional AIaaS services based on their own experimental requirements and computational resources.
\vspace{-3mm}
\section{Safety and Ethical Discussion}
All services used in the collaborative AIaaS composition dataset were collected from publicly available AIaaS platforms and service repositories. The evaluation datasets were obtained from widely used public benchmark repositories based on their documented specifications. The collaborative AIaaS composition dataset only captures observable service characteristics, evaluation behavior, and advertised specifications, without extracting publisher identities, private user information, or sensitive metadata. Furthermore, the reported evaluations are intended solely for benchmarking and comparative analysis within the dataset environment, rather than making definitive judgments about the overall quality or reliability of individual AI services.

\section{Conclusion}
In this paper, we introduced the collaborative AIaaS composition dataset, a large-scale benchmark specifically designed to support collaborative AIaaS composition research. Unlike existing datasets that are designed for traditional web services and lack AIaaS-specific attributes, the collaborative AIaaS composition dataset comprises 25,900 AIaaS services collected from five major providers across 12 AI task families, making it the largest known dataset of its kind to date. To enable systematic and fair evaluation, we generated 10,000 unique collaborative service composition requirements encompassing diverse objective functions, QoS constraints, preference weights, and corresponding high-quality composition solutions derived using a MAB-based composability algorithm. Experimental results shows that the dataset effectively differentiates the performance of various composition methods across different composition lengths and scenarios, demonstrating its suitability as a rigorous benchmarking resource. Beyond service composition, the collaborative AIaaS composition dataset broadly supports additional research directions including service recommendation, service classification, workflow orchestration, and QoS prediction, highlighting its wide practical utility. This dataset serves as a foundational resource that accelerates progress in AIaaS composition research and drives the development of more effective, scalable, and intelligent composition solutions.

\subsubsection*{Acknowledgments.}
The authors would like to thank Joshua Boland for his valuable contribution to the development and implementation of this work.

\subsubsection*{Disclosure of Interests.}
The authors have no competing interests to declare that are relevant to the content of this article.

\bibliographystyle{splncs04-unsrt}
\bibliography{ref}
\end{document}